\documentclass[
aip,
rsi,
reprint,
amsmath,
numerical,
amssymb,
]{revtex4-1}

\usepackage{graphicx}
\usepackage{dcolumn}
\usepackage{bm}
\usepackage[draft]{hyperref}
\usepackage{setspace}
\usepackage{indentfirst}
\usepackage{gensymb}
\usepackage{caption}
\usepackage[labelformat=simple]{subcaption}
\usepackage{placeins}
\usepackage{epstopdf}
\usepackage{natbib}

\newcommand{\tref}[1]{Figure \ref{#1}}

\newcommand{\mref}[2]{Figure \ref{#1}(#2)}

\begin{document}

\title{A reference-beam autocollimator with nanoradian sensitivity from mHz to kHz and dynamic range of 10$^{7}$}

\author{T. B. Arp}
\email{arpt@uw.edu, Author to whom correspondence should be addressed.}
\affiliation{Physics Department, University of Washington, Seattle, Washington, 98195, USA}

\author{C. A. Hagedorn}
\affiliation{Physics Department, University of Washington, Seattle, Washington, 98195, USA}

\author{S. Schlamminger}
\thanks{Now at National Institute of Standards and Technology, Gaithersburg, Maryland, 20899, USA}
\affiliation{Physics Department, University of Washington, Seattle, Washington, 98195, USA}

\author{J. H. Gundlach}
\affiliation{Physics Department, University of Washington, Seattle, Washington, 98195, USA}

\date{\today}

\begin{abstract}
	
	We describe an autocollimating optical angle sensor with a dynamic range of 9~mrad and nrad/$\sqrt{\textrm{Hz}}$ sensitivity at frequencies from 5~mHz to 3~kHz. This work improves the standard multi-slit autocollimator design by adding two optical components, a reference mirror and a condensing lens. This autocollimator makes a differential measurement between a reference mirror and a target mirror, suppressing common-mode noise sources. The condensing lens reduces optical aberrations, increases intensity, and improves image quality. To further improve the stability of the device at low frequencies the body of the autocollimator is designed to reduce temperature variations and their effects. A new data processing technique was developed in order to suppress the effects of imperfections in the CCD.

\end{abstract}	

\maketitle

\section{Introduction}

Optical levers have been used for non-contact measurement of angular deflection since 1826.\cite{Jones_1961} Autocollimating optical levers, or autocollimators, are insensitive to translations of the target mirror. Autocollimators are used in many experiments to provide high-resolution angle measurements and are especially useful in torsion-balance measurements of small forces.

An autocollimator collimates light into a parallel beam and then reuses the same optics to measure the deflection of a reflected beam. A collimated beam of light is produced and reflected off a mirror, set at the angle to be measured with respect to the optical axis. The reflected light acquires twice the angle of interest with respect to the incident light and is focused onto a sensor, where the position of the light spot is measured.

\begin{figure}
\centering
\includegraphics{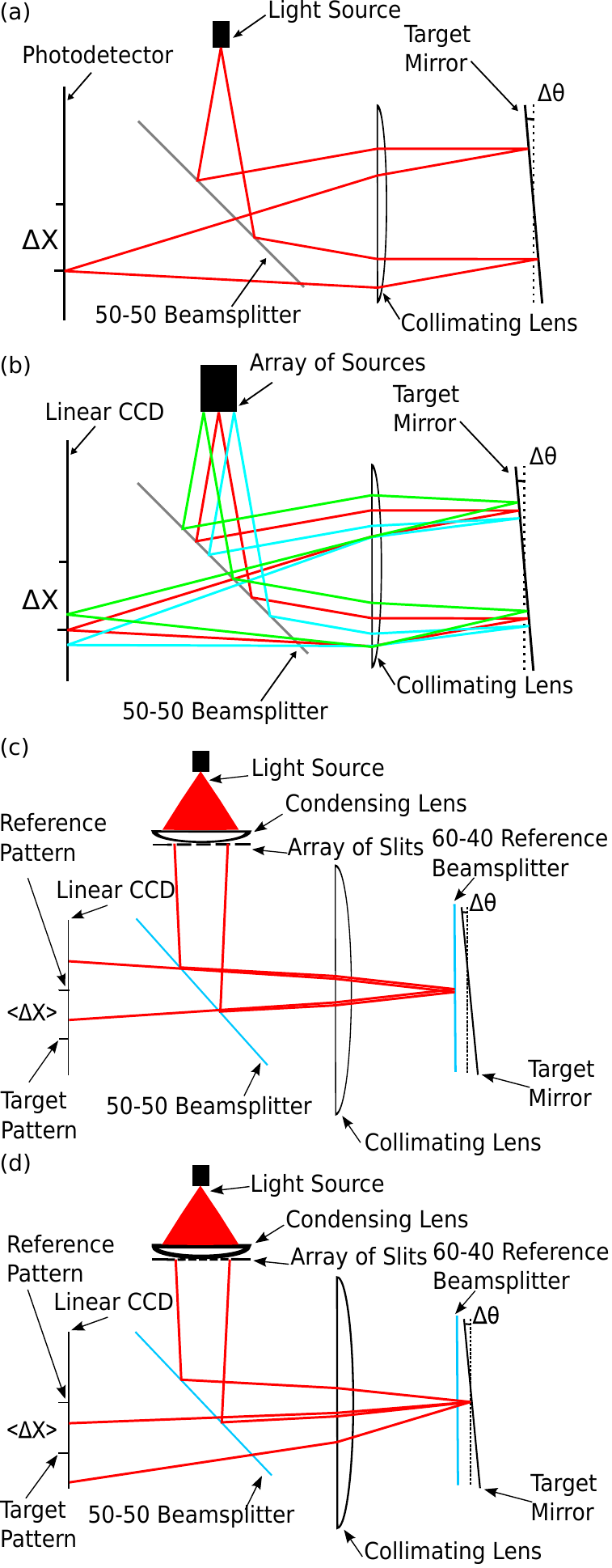}
\caption{
Schematic diagrams of autocollimator designs. (a) Standard autocollimator (b) Standard multi-slit autocollimator. Additional sources shown as blue and green rays. (c) E\"{o}t-Wash multi-slit autocollimator. Reflection from reference mirror shown. (d) E\"{o}t-Wash multi-slit autocollimator. Reflection from target mirror shown.
}
\label{fig:autocollimators}
\end{figure}

The standard autocollimator design produces light, often with a laser, that is reflected off a 50-50 beamsplitter, collimated by a lens and reflected off the target mirror [see \mref{fig:autocollimators}{a}]. The reflected column of light travels back through the collimating optics, passes through the beamsplitter, and is focused onto a position sensitive detector. An angular displacement of the mirror, $\Delta \theta$, results in a displacement of the light spot on the sensor, $\Delta X$, which is related to $\Delta \theta$ by: 
\begin{equation}
	\Delta X \approx 2 f \Delta \theta
\end{equation}
where $f$ is the focal length of the collimating lens. Variations on the basic autocollimator design have reached sensitivities of $\sim$4 nrad/$\sqrt{\textrm{Hz}}$ with dynamic ranges of $10^{4}$-$10^{5}$.\cite{SE_Pollack_2008}$^{,}$\cite{GL_Smith_1999}

Cowsik \textit{et al.} made a multi-slit autocollimator (MSA) with higher sensitivity and significantly larger dynamic range.\cite{Cowsik_2007} Cowsik used an illuminated array of slits, which are projected via the target mirror onto a linear CCD [see \mref{fig:autocollimators}{b}]. An MSA can be thought of as many standard autocollimators in parallel, increasing sensitivity by decreasing the uncertainty of the mean. 

This paper presents the design and performance of a MSA with key additions to Cowsik's design. This autocollimator was originally developed by the E\"{o}t-Wash gravity group for tests of LISA spacecraft that required nanoradian sensitivity, with one degree of dynamic range, and the capability to operate up to one meter away from the target. This autocollimator design is presently being used in torsion balance research and in a tilt sensor being developed for the Advanced LIGO project. The tilt sensor requires an autocollimator with nanoradian sensitivity in the 10~mHz to 1~Hz frequency range, specifications ideally suited for our device.

\section{Design Considerations}
	
Our MSA design [see \mref{fig:autocollimators}{c} and \mref{fig:autocollimators}{d}] adds two optical components to the MSA design: a condensing lens and a reference mirror.

\subsection{Condensing Lens} 
\label{sec:condensing}

Borrowing a technique used to reduce aberrations in projection optics, we added a condensing lens between the light source and the array of slits, immediately next to the slits.\cite{Jones_1951, Jones_1959, Habell_1948, Martin_1950} This lens redirects divergent light rays from the source into rays that hit the target mirror. This causes light to pass through the center area of the collimating lens, where the lens has fewer defects. Furthermore, the condensing lens prepares the light from each slit into a nearly parallel beam reducing the effect of aberrations of the main lens. The condensing lens reduces the effect of chromatic and spherical aberrations from the collimating lens.\cite{Malacara_2001_book} The autocollimating property of the main lens is preserved by positioning it one focal length from the slits and CCD.  The condensing lens also increases the light intensity of the pattern at the CCD, since a larger solid angle is directed along the optical path. This allows for the use of the whole intensity range of the CCD with less-intense and higher-quality fiber-coupled light sources. Further, the condensing lens ensures a uniform pattern on the CCD by redirecting all the light from each uniformly illuminated slit onto the CCD.\cite{Driscoll_1978} The source-to-slit distance is set such that the source is imaged onto the target mirror. As in the work of Jones \textit{et al.} this enforces a symmetry between the source and detector optical paths.\cite{Jones_1951} To first order, this symmetry is not required due to the autocollimating property of the objective lens, but symmetric alignment yields maximum image intensity and further suppresses aberration.

\subsection{Reference Mirror}

Our second optical addition to the MSA design is a partially-silvered mirror in front of the target mirror which acts as a reference. Another reference beam angle sensor arrangement was recently reported by Li \textit{et al.}\cite{Li_2013} The reference mirror returns 40\% of the beam through the optics, and defines an angular position against which the target mirror can be compared. The rest of the beam goes through the reference mirror and is reflected by the target mirror. Two images of the slits form on the CCD, the first reflected from the reference and the second from the target mirror. The displacement between the two patterns measures the angle between the reference mirror and the target mirror. This differential  measurement removes noise that is common-mode to both the target and reference mirrors and makes most potential noise sources associated with the autocollimator itself second-order. For example, the effects of thermal expansion of the autocollimator body, intensity drift, and vibration are all reduced by using the reference mirror. The light returning from the target mirror is attenuated twice by the reference beamsplitter, therefore a 60T-40R mirror is used to match the intensity of the light from the reference mirror and the target mirror at the detector (the optimal ratio is 0.618T-0.382R). Some reflected light from the target mirror is reflected again on the trip back through the reference mirror. This reflects off the target mirror again and is at twice the angle with respect to the reference and is therefore either distinct from the target pattern or off the CCD.

\subsection{Thermal Considerations}

For millihertz stability the autocollimator body was made from thick aluminum, giving it long thermal time constants and high thermal conductivity. Remaining effects of thermal gradients over the body of the autocollimator, while suppressed to first order by the differential measurement, could introduce low frequency noise due to asymmetric thermal expansion. Therefore, the autocollimator is designed to have no internal heat sources except for the CCD chip; all of the CCD readout electronics are thermally separated from the autocollimator body and light is brought in through an optical fiber. 

The reference beam suppresses many common-mode effects. However, at least two effects are not canceled by the reference beam: the thermal expansion of the CCD and the effect of variations in the wavelength of the light source. Thermal expansion of the linear CCD causes a false change in angle as the silicon CCD chip expands but the patterns remain in the same position. In the worst case, a temperature variation of 0.1~K would result in a false change in angle of 8 nrad. Therefore, the CCD is heat-sunk to the body of the autocollimator to push the CCD's thermal time constant into the millihertz. Chromatic aberrations in the lenses can create a sensitivity to variations in the wavelength of the light source which is not canceled by the reference. Therefore, the LED light source is heat-sunk to a large insulated thermal mass to give it a thermal time constant $>$5000~s.

\section{Device Implementation}

\begin{figure}
\centering
\includegraphics{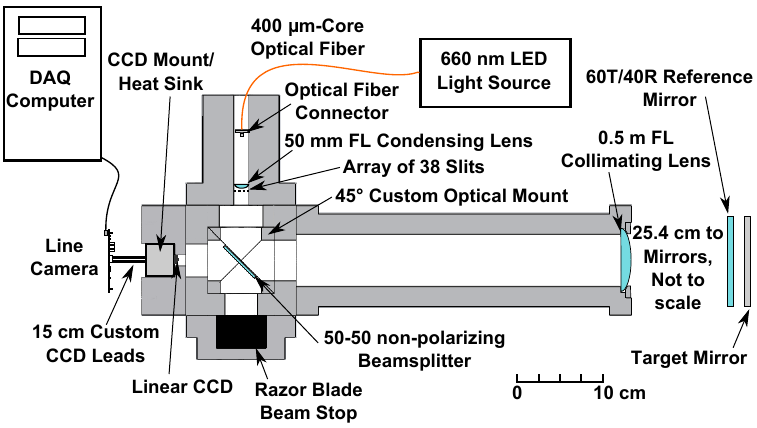}
\caption{Cross sectional view of our multi-slit autocollimator. The device measures the angle between the reference mirror and target mirror about the vertical axis. The autocollimator body is to scale.}
\label{fig:autocollimatorDevice}
\end{figure}

\begin{figure}
\centering
\includegraphics{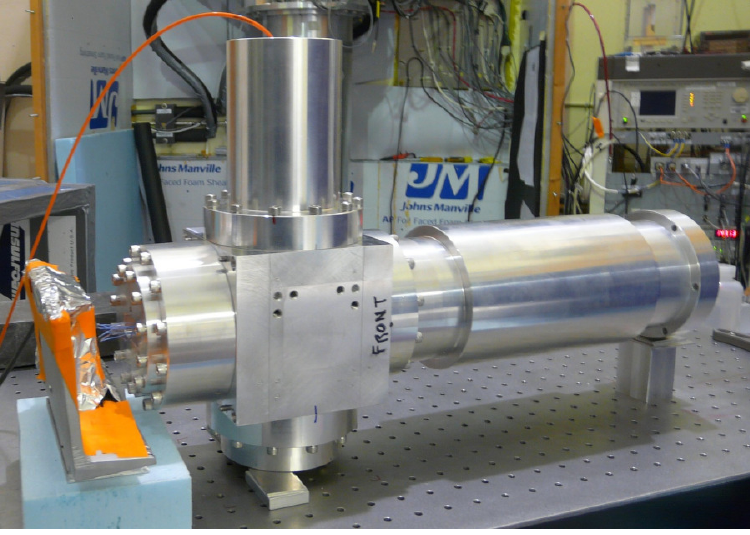}
\caption{The multi-slit autocollimator, during in-air noise runs. Foam shielding, used to stabilize temperature and eliminate air currents, removed for picture.}
\label{fig:autocollimatorPicture}
\end{figure}

\begin{figure}
\centering
\includegraphics{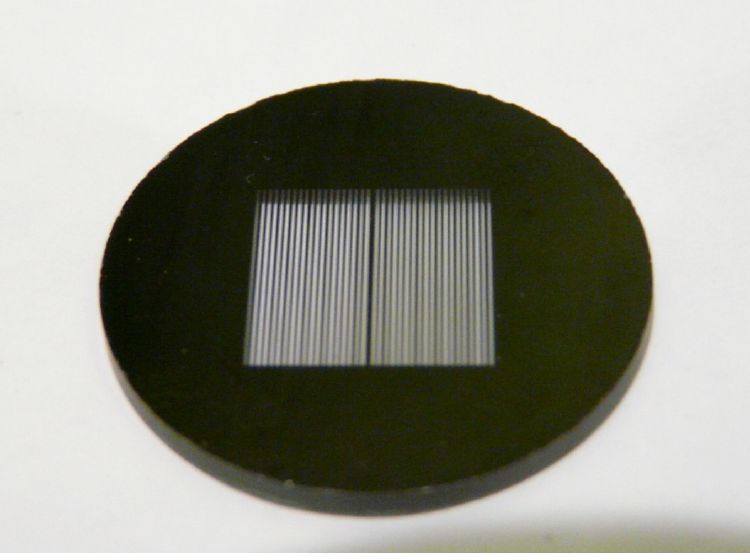}
\caption{The array of slits used in the autocollimator}
\label{fig:grating}
\end{figure}

Our multi-slit autocollimator (see \tref{fig:autocollimatorDevice} and \tref{fig:autocollimatorPicture}) is built from extruded 6061-T651 Aluminum for stability and vacuum compatibility. To avoid laser speckle we use a 660~nm Superbright LED (ThorLabs: M660F1) as a light source, coupled into a multimode optical fiber with a 400~$\mu$m core diameter and 0.38 NA (ThorLabs: M28L02). We use an array of 38 slits, with a missing peak in the center, made using a chrome photomask on quartz glass [see \tref{fig:grating}]. Each slit is 145~$\mu$m wide and the spacing between the slits is 145~$\mu$m. All optics, except for the photomask, are anti-reflection coated to reduce reflections which might interfere with the final image. A beamstop made from stacked razor blades is installed opposite the light source to absorb light transmitted through the 50-50 beamsplitter.\cite{Hecht_2002} Fixed target noise tests were done with the reference and target mirrors held against a 1.2~cm thick solid square Invar frame that inclines the target mirror by 1$\degree$ with respect to the reference mirror.

For autocollimation, the slits are positioned such that the path length from the slits to the collimating lens is one focal length. Similarly, the CCD is positioned at one focal length from the collimating lens; the CCD holder was shimmed for small position adjustments. The autocollimator is read out with a Mightex TCN-1209-U line camera modified to separate the CCD from the readout electronics. The camera uses a 28.6~mm long Toshiba 1209DG 12-bit linear CCD with 2048 $14 ~ \mu \textrm{m}$ square pixels operated at a rate of 3300 frames/second. A high frame rate allows for differential subtraction of common-mode high-frequency noise and prevents aliasing. 

\begin{figure*}
\centering
\includegraphics{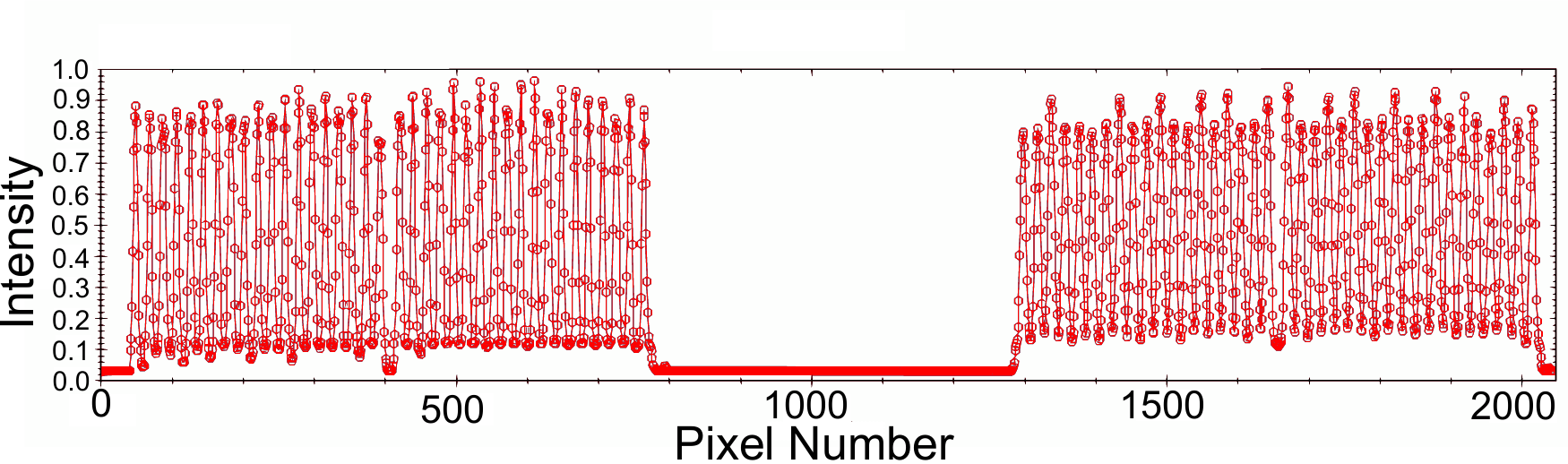}
\caption{A frame from the CCD. Frames are processed at 3330 frames/s. The left pattern is from the reference mirror and the right pattern is from the target mirror.}
\label{fig:fringes}
\end{figure*}%

\section{Image Processing}

\begin{figure}        
\centering
\includegraphics{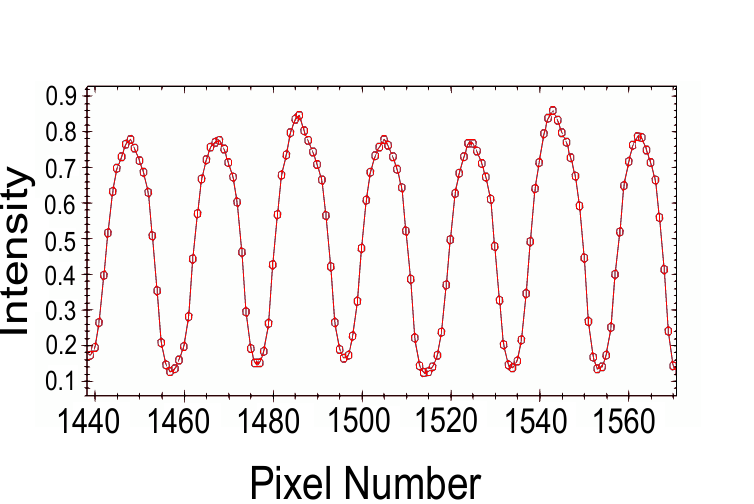}
\caption{Close-up of peaks in the pattern on the CCD.}
\label{fig:peaks}
\end{figure}

The patterns reflected from the reference and target mirrors are imaged onto the linear CCD and used to read out the autocollimator. The patterns are images of the array of slits [see \tref{fig:fringes}]. The technique used in the MSA reported by Cowsik \textit{et al.} and early versions of our software was to find the centroid of each peak and average the centroid locations to calculate the position of each pattern.\cite{Cowsik_2007} This method has the advantage of being simple and works well for many applications, however, there are disadvantages for high sensitivity and low frequency measurements. The CCD pixels were measured to be both nonlinear and noisy at low intensity. A centroid algorithm weights all pixels equally, therefore, increased noise in the low intensity pixels obscures the better sensitivity available from the higher intensity pixels. In addition, our centroid algorithm used thresholds to identify peaks and was therefore sensitive to the particular choices of threshold values and to intensity fluctuations that cause pixels to cross the threshold, introducing low frequency noise. We developed a new data processing algorithm to reduce these effects. Our algorithm identifies peaks by finding local maxima and fits each peak with a Gaussian function [see \tref{fig:peaks}]. This algorithm takes the logarithm of each peak and fits a quadratic function with weighted ordinary least squares. 

We perform this fit using a custom variation of the ordinary least squares routine from the GNU Scientific Library (function name: \texttt{gsl\textunderscore multifit\textunderscore wlinear}) which we optimized for speed to give a single peak fit rate faster than 253~kHz.\cite{GNU_GSL} The only fit parameter that matters in this application is the center of the Gaussian, therefore the weights can be tuned to emphasize center detection. We weight the logarithm of the data points with a Gaussian of fixed width approximately centered on each peak. This emphasizes the tops and upper sides of the peak over the lower portions and produces the best performance of all weighting schemes explored. This algorithm has lower noise compared to the centroid algorithm as the weighting scheme de-weights the noisy low intensity pixels, is not sensitive to a threshold, and reduces sensitivity to intensity variation.

The autocollimator has a reference pattern, therefore each peak on the target pattern can be compared to the corresponding peak on the reference pattern allowing the autocollimator to function as multiple differential autocollimators in parallel. The differential measurement is made by subtracting the position of each peak on the reference pattern from the corresponding peak on the target pattern. The angle is determined by the average peak separation.

\FloatBarrier

\section{Alignment and Distortion}

Imperfections can cause the image of the slits on the CCD [see \tref{fig:fringes}] to be distorted. If the CCD is not positioned at the focal length of the main lens the peaks will be blurred. If either the reference mirror or target mirror are curved, then they will act as a lens and change the focal distance for the light reflected from that mirror. For example, if there are different stresses on the two mirrors (e.g. one mirror mount is tighter) it can manifest as a slight asymmetry between the patterns on the CCD, as the pattern from one mirror will be better resolved than the other. This asymmetry can be minimized by fine tuning the position of the CCD to sit between the two focal spots. In addition to distortion of the whole pattern, ``double peaking" can occur where the peaks in one pattern have two distinct maxima. 

When properly aligned, the light source is imaged onto the target mirror. In addition, diffraction from the slits causes a diffraction pattern at the mirrors. Therefore, the image on the target mirror is the diffraction pattern convolved with the image of the light source. Care must be taken not to asymmetrically clip the diffraction pattern, as this will cause image distortion.

Distortion can be minimized through careful alignment of the device. A robust alignment strategy is as follows. First, the slits are positioned one focal length away from the main lens to guarantee collimation. Next, the target and reference mirrors are adjusted until the image of the source/diffraction pattern is centered on them. Then, the position of the light source is adjusted until the image of the light source (convolved with the diffraction pattern) is sharp on the target mirror (see Section \ref{sec:condensing}). Finally, the position of the CCD is adjusted until the images are in sharp focus.

\section{Results and Discussion}

\begin{figure}
\centering
\includegraphics{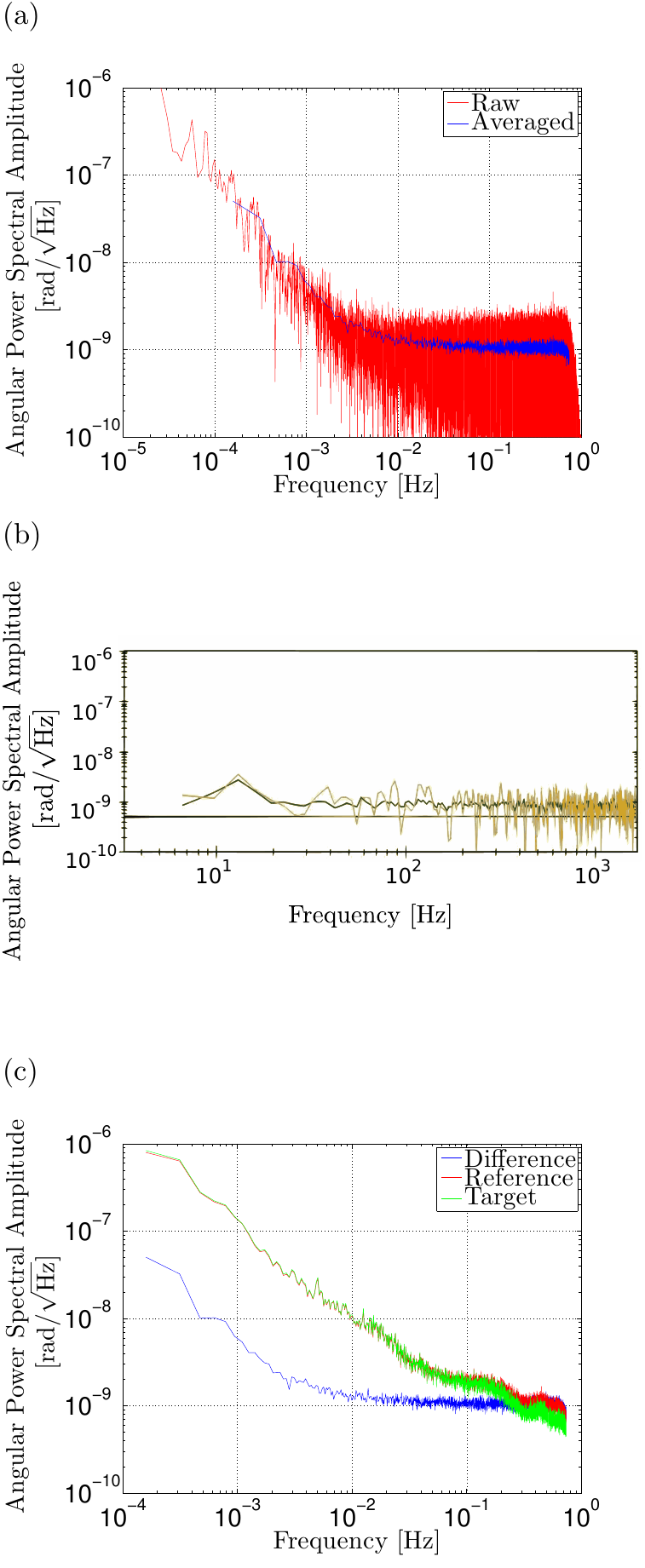}
\caption{
Sensitivity of the autocollimator for Invar-mounted mirror pair. (a) Power Spectral Amplitude of the difference signal at low frequencies. (b) High frequency sensitivity. Yellow is the full spectrum, green is the averaged spectrum and the black line is 0.5 nrad/$\sqrt{\textrm{Hz}}$. (c) Averaged sensitivity of the target mirror, reference mirror, and the difference between the mirrors.
}
\label{fig:spectra}
\end{figure}%

\begin{figure}
\centering
\includegraphics{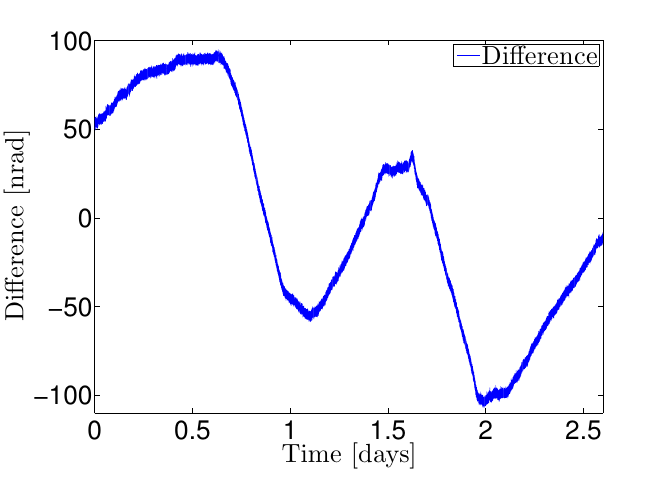}
\caption{Difference angle data for the noise run shown in \tref{fig:spectra}. Daily variation is believed to be due to 0.25-0.5~K daily variation in lab temperature.}
\label{fig:dataTrace}
\end{figure}%

Noise tests were initially carried out with the autocollimator operating in air. However, the autocollimator proved to be sensitive to small variations in air pressure, such as the door to the lab opening and closing, due to slight bowing of the target and reference mirrors. In addition, the autocollimator is sensitive to air currents along the beampath. Most applications have both the target and reference mirrors in vacuum, eliminating air sensitivity. Foam baffles placed around the external beampath reduced, but could not eliminate, this effect. For final testing the autocollimator was placed in a vacuum chamber pumped down to moderate vacuum ($\sim$1~kPa), maintaining gas thermal conductivity but eliminating pressure fluctuations. In vacuum, the differential autocollimator sensitivity improved by approximately a factor of three at frequencies below 50~mHz compared with the best performance at atmospheric pressure. When attached to a beam balance for the dynamic tests discussed below, the autocollimator body is in air but the reference mirrors, target mirror, and most of the external beampath is in vacuum. The high frequency noise floor on the beam balance is comparable to the noise measured with the whole autocollimator in vacuum.

To measure the noise introduced by thermal effects, the body of the autocollimator and the CCD were instrumented with temperature sensors. The leading effects of temperature variation were thermal expansion of the CCD, changing the apparent displacement of the patterns, and temperature variation in the LED, causing fluctuations in the wavelength of the light source. Adding heat sinks to the CCD and the LED, increasing their thermal time constants to $>$1000~s, was sufficient to reduce the effect of temperature variation to below the $1$~nrad/$\sqrt{\textrm{Hz}}$ noise floor from 10~mHz to 1~Hz.

\mref{fig:spectra}{a} shows the sensitivity of the autocollimator at low frequencies and \mref{fig:spectra}{b} shows the sensitivity at high frequencies. The noise between 1~Hz and 10~Hz is consistent with the levels seen in \mref{fig:spectra}{a} and \mref{fig:spectra}{b}. The autocollimator has $\sim$1~nrad/$\sqrt{\textrm{Hz}}$ sensitivity at all frequencies down to $\sim$5~mHz where the noise begins to rise as 1/$f$ in amplitude. \mref{fig:spectra}{c} shows the sensitivity of the reference and target patterns compared to the difference signal.  As expected, at high frequencies the difference noise floor is inherently a factor $\sqrt{2}$ higher than the noise floors of the reference and target patterns due to subtraction. The differential measurement effectively suppresses noise at frequencies below $\sim$0.2~Hz by an order of magnitude. As hinted by \mref{fig:spectra}{c}, alignment adjustments can give single-mirror noise floors as low as 0.5~nrad/$\sqrt{\textrm{Hz}}$, with a differential noise floor of 0.7~nrad/$\sqrt{\textrm{Hz}}$. When the CCD was immobilized for low-frequency tests presented here, the reference signal was noisier than the target signal, limiting differential performance to 1~nrad/$\sqrt{\textrm{Hz}}$.

After internal optical alignment, our autocollimator is robust and practical. It operates in both single-mirror and reference-mirror modes with ease. It can be removed from an apparatus and reattached/aligned in minutes. The autocollimator operates in ambient room light with a slightly elevated noise floor. External connections to the instrument are a USB cable and an optical fiber; no analog electrical or ground connections are required. The instrument's optimal working point was adjusted for a $25$~cm separation between the objective and the mirrors; but similar performance can be achieved for working distances between 10-40~cm without realignment. The aperture of the main lens restricts the maximum working distance; our first prototype MSA with a 15.25~cm objective lens operated with a $\sim$1~nrad/$\sqrt{\textrm{Hz}}$ differential noise floor at frequencies above 0.1~Hz at a working distance of $1$~m. The dynamic measurements discussed below were made with a 7 cm separation in optical path length between the reference and target mirrors.

\begin{figure}
\centering
\includegraphics{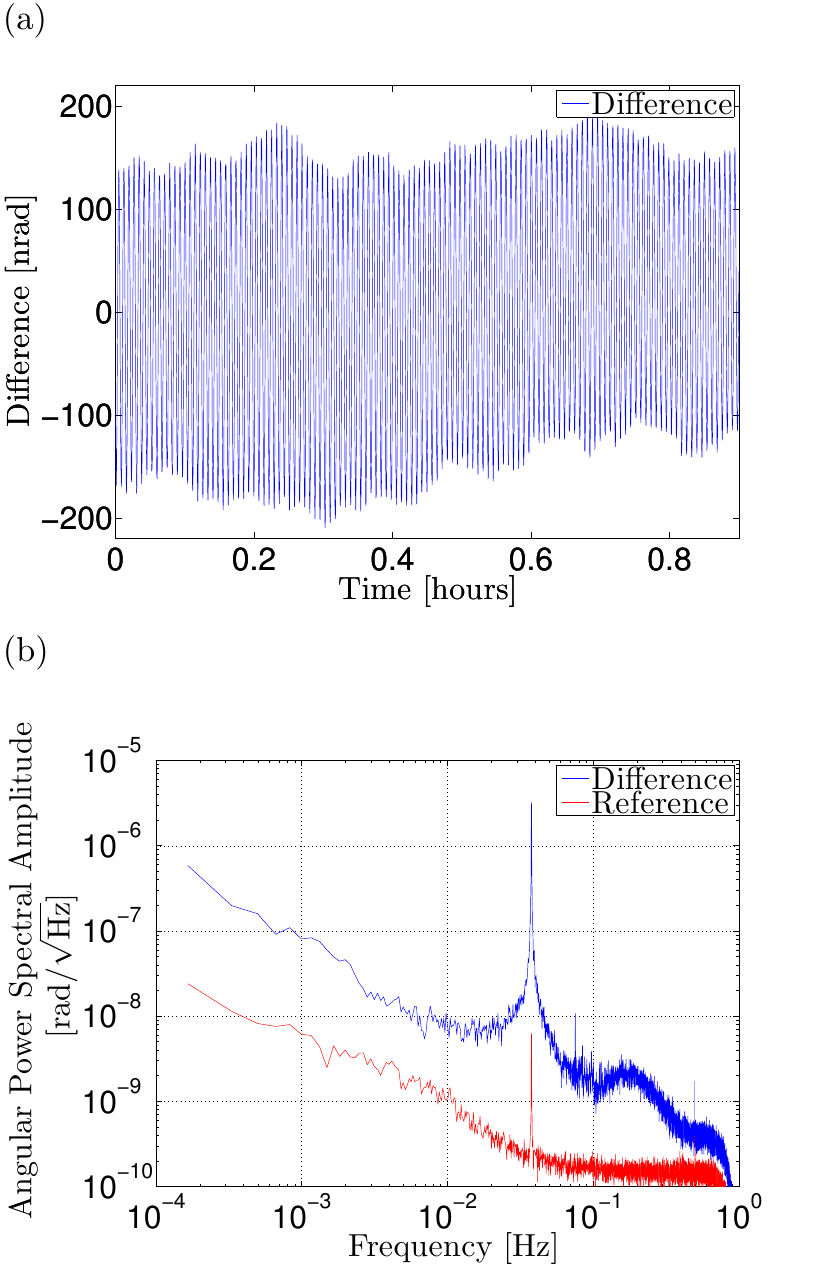}
\caption{
MSA performance on the beam balance with four-bounce signal amplification. (a) Short section of autocollimator data from tiltmeter beam balance. (b) Ground tilt measured with the autocollimator on a beam balance.
}
\label{fig:tiltMeasure}
\end{figure}%

To demonstrate the dynamic performance of the autocollimator it was mounted on a beam balance apparatus. The beam balance consists of a beam with a large moment of inertia suspended from a pair of flexures very close to its center of mass, making a harmonic oscillator with a Q $>$ 5000 and resonant period of 26~s. To further increase the angle sensitivity, light is reflected four times off the beam amplifying the signal and lowering the effective noise floor by a factor of 4. In this configuration the autocollimator measures this balance with $\sim$250 prad/$\sqrt{\textrm{Hz}}$ sensitivity above 50~mHz, as shown in \mref{fig:tiltMeasure}{b}.

\begin{figure}
\centering
\includegraphics{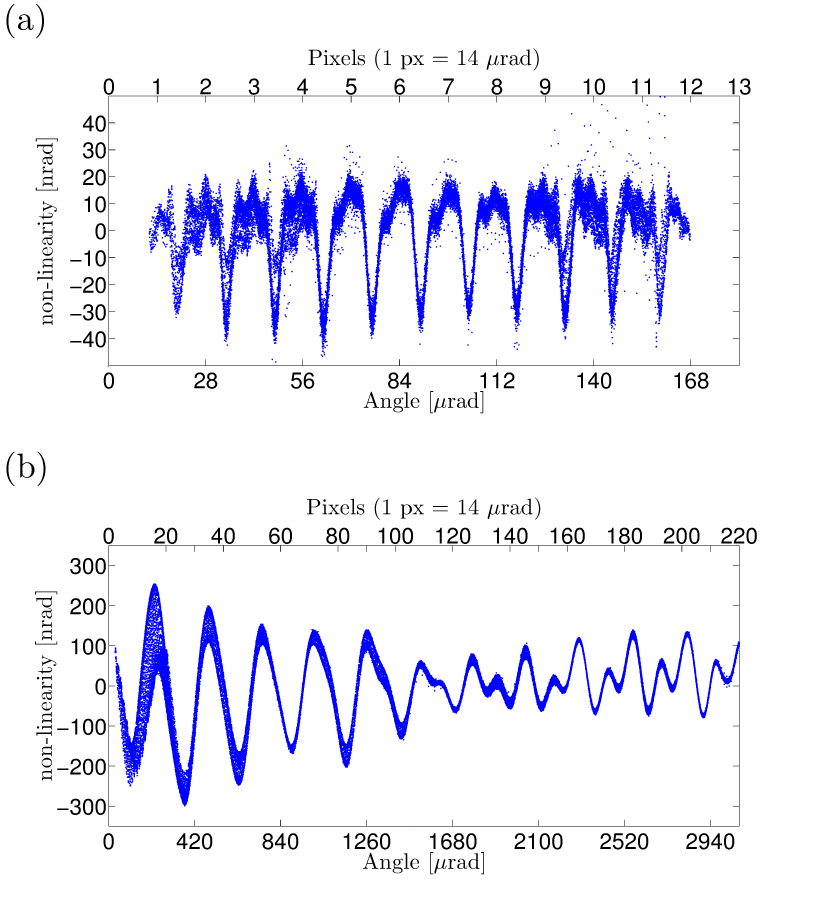}
\caption{
MSA non-linearity for (a) small amplitude and (b) large amplitude.
}
\label{fig:tilt}
\end{figure}

The autocollimator's dynamic range is measured to be 9.3$\pm$0.1~mrad (0.53$\pm$0.006$\degree$).  With noise of $\sim$1 nrad the autocollimator has a signal-to-noise ratio of $\sim$10$^{7}$. Tests on the beam balance allowed us to measure the linearity of the autocollimator by assuming Hooke's Law for small oscillations ($\theta \ll \pi$) and examining the difference between the measured beam balance motion and linear-phase finite-impulse-response (FIR) low-passed beam balance motion. There are two important scales for non-linearity in this instrument, pixel-level and large amplitude. \mref{fig:tilt}{a} shows the non-linearity for small oscillations. As the target pattern moves by a pixel (1 pixel = 14~$\mu$rad) there is a non-linearity of approximately 60~nrad which gives a fractional non-linearity of $<$0.5\%. \mref{fig:tilt}{b} shows the non-linearity for large oscillations, as the patterns move by about 250~$\mu$rad there is at most a non-linearity of 500~nrad which results in a fractional non-linearity of at most 0.2\%. Over the measured 3~mrad range on the beam balance the non-linearity is smaller than 500~nrad, which gives a conservative full scale non-linearity of $1.7 \times 10^{-4}$.

One of the limiting factors for multi-slit autocollimator sensitivity is the number of slits used. The noise of a multi-slit autocollimator improves as the square root of the number of slits. We used only 38 slits in order to have room on the CCD for 10~mrad of dynamic range. The multi-slit autocollimator by Cowsik \textit{et al.} used 110 slits and a focal length of 1~m.\cite{Cowsik_2007} If we were to use 82 slits (the maximum we can fit on the CCD with our slit size) and a 1~m focal length our noise floor would be 200 prad/$\sqrt{\textrm{Hz}}$ at high frequency with no differencing or multiple reflection amplification. A longer linear CCD would increase the number of slits we could use while maintaining the same dynamic range. A linear CCD only uses a line from the image of the slits, but advances in CCD technology and signal processing may make it possible to read out a two dimensional CCD at a high enough frame rate, further increasing sensitivity.

In conclusion, the multi-slit autocollimator design is an improvement to angle-measuring technology. By employing a reference mirror a MSA can measure the angle differentially, which suppresses multiple noise sources. A condensing lens reduces the effect of optical aberrations and increases intensity throughput. Challenges of the MSA design include controlling the distortion of the patterns and processing the image from the CCD. With the condensing lens, reference mirror, and other improvements the multi-slit autocollimator design was used to build a sensor with 1~nrad/$\sqrt{\textrm{Hz}}$ sensitivity over a frequency range of 5~mHz to 3~kHz, with 9~mrad of dynamic range and a linearity better than 0.02\%.


\section{Acknowledgments}

We would like to thank Larry Stark for his artistry in the University of Washington machine shop. We thank the LISA and LIGO collaborations, along with NASA (Grant: NNX08AY66G) and the NSF (Grants: PHY0653863, PHY0969199 and PHY0969488) for funding the development of the autocollimator. We thank Jenna Walrath who worked on this project for her Research Experience for Undergraduates (REU) program. We thank Matt Turner, Krishna Venkateswara, and the E\"{o}t-Wash Group for help and advice. We thank the Center for Experimental Nuclear Physics and Astrophysics (CENPA) for use of its facilities.


%

\end{document}